\documentclass[twocolumn,aps,prl,groupedaddress]{revtex4}

\usepackage{epsfig,amsfonts,amsmath}
\usepackage{graphicx}

\newcommand{\be}{\begin{equation}}
\newcommand{\ee}{\end{equation}}
\newcommand{\ba}{\begin{eqnarray}}
\newcommand{\ea}{\end{eqnarray}}

\def\M#1#2{M_{#1}\left( #2\right) }
\def\R#1#2{R_{#1}\left( #2 \right) }
\def\Rn#1#2#3{R_{#2}^{[#1]}\left( #3 \right) }
\def\ceil#1{\left\lceil #1 \right\rceil}
\def\floor#1{\left\lfloor #1 \right\rfloor}
\def\e{\epsilon}

\def\bea{\begin{eqnarray}}
\def\eea{\end{eqnarray}}
\def\ben{\begin{eqnarray*}}
\def\een{\end{eqnarray*}}

\def\non{\nonumber}
\def\>{\rangle}
\def\<{\langle}

\newcommand{\fig}[1]{Fig.~\ref{fig:#1}}

\DeclareMathOperator{\tr}{Tr}

\begin{document}


\title{Arbitrarily accurate composite pulse sequences}

\author{Kenneth R. Brown$^\dagger$, Aram W. Harrow$^\dagger$, 
		and Isaac L. Chuang$^{\dagger*}$}




\affiliation{$^\dagger$Center for Bits and Atoms, MIT, Cambridge, MA 02139
\\
$^*$Department of Physics, MIT, Cambridge, MA 02139
}

\date{\today}

\begin{abstract}
Systematic errors in quantum operations can be the dominating source
of imperfection in achieving control over quantum systems.  This
problem, which has been well studied in nuclear magnetic resonance,
can be addressed by replacing single operations with composite
sequences of pulsed operations, which cause errors to cancel by
symmetry.  Remarkably, this can be achieved without knowledge of the
amount of error $\epsilon$.  Independent of the initial state of the
system, current techniques allow the error to be reduced to
$O(\e^3)$.  Here, we extend the composite pulse technique to cancel
errors to $O(\e^n)$, for arbitrary $n$.
\end{abstract}


\maketitle


Precise and complete control over closed quantum systems is a
long-sought goal in atomic physics, molecular chemistry, condensed
matter research, with fundamental implications for
metrology\cite{freqmet,Bennett98a} and
computation\cite{mikeandike,Steane98a}.  Achieving this goal will
require careful compensation for errors of both random and systematic
nature.  And while recent advances in quantum error
correction\cite{qec,Steane96a,Knill97a} allow all such errors to be
removed in principle, active error correction requires expanding the
size of the quantum system, and feedback measurements which may be
unavailable.  Furthermore, in many systems, errors may be dominated by
those of systematic nature, rather than random errors, as when the
classical control apparatus is miscalibrated or suffers from
inhomogeneities over the spatial extent of the target quantum system.


Of course, systematic errors can be
reduced simply by calibration, but that is often impractical,
especially when controlling large systems, or when the required
control error magnitude is smaller than that easily measurable.
Interestingly, however, systematic errors in controlling quantum
systems can be compensated without specific knowledge of the magnitude
of the error.  
This fact is lore\cite{Freeman-book} in the art of NMR,
and is achieved using the method of {\em composite pulses}, in which a
single imperfect pulse with fractional error $\e$ is replaced with a
sequence of pulses, which reduces the error to $O(\e^n)$.

Composite pulse sequences have been constructed to correct for a wide
variety of systematic errors \cite{Levitt79,Tycko83a,Freeman-book}.  These include pulse
amplitude, phase, and frequency errors and can be applied to any system with sufficient control.  As system control increases, new uses for composite pulses emerge.
A remarkable example is the recent teleporation of an atomic state in ion traps \cite{Reibe04,Barrett04}. Barret {\it et al.} use a composite pulse for individual addressing, while Reibe {\it et al.} use a composite pulse to perform two qubit operations.

In the context of spectroscopy,
the goal is often to maximize the measurable signal from a 
system which
starts in a specific state.  Thus, while composite sequences have been
developed\cite{Levitt83} which can reduce errors to $O(\e^n)$ for
arbitrary $n$, these sequences are not general and do not apply, for
example, to quantum computation, where the initial state is arbitrary,
and multiple operations must be cascaded to obtain desired multi-qubit
transformations.

Only a few composite pulse sequences are known which are {\em fully
compensating},\cite{Levitt86a,Wimperis:94} meaning that they work on
any initial state and can replace a single pulse without further
modification of other pulses.  As has been theoretically
discussed\cite{Cummins00a,Jones02,McHugh:04} and experimentally
demonstrated in ion traps \cite{Gulde:03,Reibe04,Barrett04} and Josephson junctions \cite{Collin04}, these sequences can be valuable for
 precise
single and multiple-qubit control using gate voltages or laser
excitation.

Previously, the best fully compensating composite pulse sequence
known\cite{Wimperis:94,Cummins00a,Jones02,McHugh:04} could only
correct errors to $O(\e^3)$\endnote{ In
Ref. \cite{Cummins00a,Jones02,McHugh:04}, the distance measure used is
one minus the fidelity, $1-\| V^\dagger U\|$ (``the infidelity'')
where $\| A\|$ is the norm of $A$.  We use instead the trace distance,
$\| V-U\|$, following the NMR community.  Thus, our composite pulses
which are $n$th order in trace distance are $2n$th order in
infidelity.}.  Here, we present a new, and systematic technique for
creating composite pulse sequences to correct errors to $O(\e^n)$, for
arbitrary $n$.  The technique presented is very general and can be used to correct a wide variety of systematic errors. Below, our technique is illustrated for the specific
case of systematic amplitude errors, using two approaches. Also
discussed is the number of pulses required as a function of $n$.

The problem of systematic amplitude errors is modeled by representing 
single qubit rotations as
\be
  \R{\phi}{\theta} = \exp \left[ -i \frac{\theta}{2} \sigma_\phi \right]
\,,
\ee
where $\theta$ is the desired rotation angle about about the axis
 that makes the angle $\phi$ with the $\hat{x}$-axis and lies in the
 $\hat{x}-\hat{y}$ plane, $\sigma_\phi=\cos (\phi) X 
+ \sin (\phi) Y$, and $X$ and $Y$ are Pauli operators.
$\R{\phi}{\theta}$ is the ideal operation, and due to errors, the
actual operation is, instead, $\M{\phi}{\theta} =
\R{\phi}{\theta(1+\e)}$, where the angle of rotation differs from the
desired $\theta$ by the factor $1+\epsilon$.  Note that $\phi$ and
$\theta$ may be specified arbitrarily, but the error $\epsilon$ is
fixed for all operations, and unknown.


\noindent
{\bf Two methods for constructing composite pulses.}  A composite
pulse sequence $\Rn{n}{\phi}{\theta}$ is a sequence of operations
$\{M_\phi(\theta)\}$ such that $\Rn{n}{\phi}{\theta} =
\R{\phi}{\theta} + O(\e^{n+1})$, for unknown error $\e$.  To construct
$\Rn{n}{\phi}{\theta}$, we begin with two simple observations: first,
$\R{\phi}{-\theta\e}\M{\phi}{\theta}=\R{\phi}{\theta}$ and second,
$\M{\phi}{2k\pi}=\pm\R{\phi}{2k\pi\e}$ when $k$ is an integer.  A
composite pulse sequence can thus be obtained by finding ways to
approximate $\R{\phi}{-\theta\e}$ by a product of operators
$\R{\phi_l}{2k_l\pi\e}$.  We obtain this using two approaches.

The first approach we call the Trotter-Suzuki (TS) method.  Suzuki has
developed a set of Trotter formulas that when given a Hamiltonian $B$
and and a series of Hamiltonians $\{A_l\}$ such that $B=\sum{A_l}$
there exists a set of real numbers $\{p_{jn}\}$ such that
\be
	\exp \left(-i B t\right)= 
	\prod_{j,l} \exp \left(-i p_{jn }A_l  t \right)+O(t^{n+1}) 
\,,
\label{TS}
\ee
and $\sum_j p_{jn}=1$ \cite{Suzuki:92}.  Without loss of generality,
we may limit ourselves to expansions where the $p_{jn}$ are rational
numbers, and assume the goal is to approximate
$\R{0}{-\theta\e}$. Using Eq.~(\ref{TS}), we set $t=\e$
and $B=-(\theta/2) X$. Then we choose $A_1=A_3= m \pi (X \cos \phi
+Y\sin \phi)$ and $A_2=2 m \pi (X \cos \phi - Y \sin \phi)$ where
$\phi$ and $m$ fulfill the conditions that $ 4 m \pi \cos \phi=\theta/2$
({\em i.e.,} $A_1+A_2+A_3=B$) and $q_{jn} = p_{jn}m$ is an
integer. This yields an $n$th order correction sequence
\ba
\nonumber
F_n&=&\prod_{j} \M{\phi}{2\pi q_{jn}}\M{-\phi}{4\pi
	q_{jn}}\M{\phi}{2\pi q_{jn}}
\\
       &=& \R{0}{-\theta\e}+O(\e^{n+1}) = \Rn{n}{0}{-\theta\e}
\ea
and the associated $n$th order composite pulse sequence
$F_n\M{0}{\theta} = \Rn{n}{0}{-\theta\e} \R{0}{\theta\e}\R{0}{\theta}
= \Rn{n}{0}{\theta}$, thus giving a composite pulse sequence of
arbitrary accuracy.

The second approach we refer to as the Solovay-Kitaev (SK) method, as
it uses elements of the proof of the Solovay-Kitaev theorem
\cite{KSV}.
First, note that rotations $U_{k}(A) = I+A\e^k+O(\e^{k+1})$ can be
constructed for arbitrary $2 \times 2$ Hermitian matrices $A$, and
$k\geq 1$, recursively.  This is done using an observation (from
\cite{KSV}) relating the commutator $[A,B] = AB-BA$ to a sequence of
operations, $\exp (-iA \e^l) \exp (-iB \e^m) \exp (iA \e^l) \exp (iB
\e^m)=\exp ([A,B] \e^{l+m})+O(\e^{l+m+1})$.  Thus to generate
$U_{k}(A)$ it suffices to generate $U_{\lceil{k/2}\rceil}(B)$ and
$U_{\lfloor{k/2}\rfloor}(C)$ such that $[B,C]=A$ (choices of integers
other than $\ceil{k/2}$ and  $\floor{k/2}$ which sum to $k$ are also
fine, but less optimal).

Next, we inductively construct a composite pulse sequence $F_n$ for
$\R{0}{\theta}$.  Note that the first order correction sequence can be
written as $F_1=\M{\phi}{2\pi}\M{-\phi}{2\pi} =
\R{0}{-\theta\e}+O(\e^2)$ by selecting $4\pi\cos(\phi)=\theta$.
Assume we have $F_n=\R{0}{-\theta\e}-i A_{n+1}\e^{n+1}+O(e^{n+2})$.
We can then construct a sequence to correct for the next order, using
$F_{n+1}=U_{n+1}(A_{n+1}) F_n$, where $U_{n+1}(A_{n+1})$ is
constructed as above.  Iteratively applying this method for
$k=1,\ldots,n$ yields an $n$th order composite pulse sequence,
$F_n\M{0}{\theta} = \Rn{n}{0}{\theta}$, for any $n$.  This method,
which appears to be unrelated to previous composite pulse
techniques\cite{Levitt83,Freeman-book}, gives an efficient algorithm
to calculate sequences for specific $\theta$ and $\phi$ but not
necessarily a short analytical description of the sequence. Furthermore, the Solovay-Kitaev technique relies on general properties of Hamiltonians and can be applied without modification to other systematic error models, {\it e.g.}, frequency errors.

\noindent{\bf Examples.}  The TS and SK techniques described above are
general and apply to a wide variety of errors; explicit application of
the techniques to generate $\Rn{n}{0}{\theta}$ sequences for specific
$n$ can take advantage of symmetry arguments, composition of
techniques, and relax some of our assumptions to minimize both the
residual error and the sequence length.

\begin{figure}
\includegraphics[width=7cm]{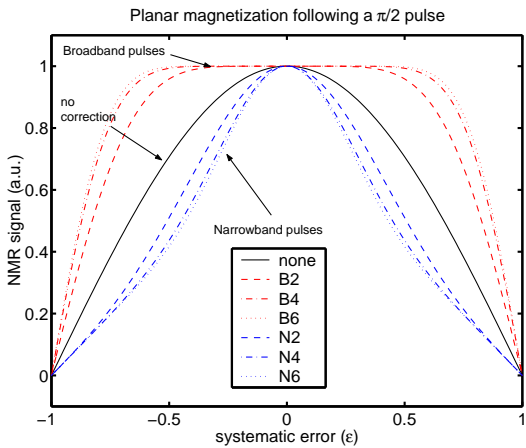}
\caption{
Comparison of the narrowband and broadband composite pulse sequences
generated by the TS method.  The Wimperis BB1, PB1, and NB1 sequences are
included in this family, and are equivalent to B2, P2, and N2.
}
\label{fig:signal}
\end{figure}

First, we explicitly write out the TS composite pulses and connect
them to the well-known pulse sequences of Wimperis \cite{Wimperis:94}.
We choose to use the TS formulas that are symmetric under reversal of
pulses, {\em i.e} an anagram.  These formulas remove all even-ordered
errors by symmetry, and thus yield only even-order composite pulse
sequences.  For convenience, we introduce the notation
$S_1(\phi_1,\phi_2,m)=\M{\phi_1}{m\pi}\M{\phi_2}{2m\pi}\M{\phi_1}{m\pi}
$ and $S_n(\phi_1,\phi_2,m)=S_{n-1}(\phi_1,\phi_2,m)^{4^{n-1}} \times
S_{n-1}(\phi_1,\phi_2,-2m)S_{n-1}(\phi_1,\phi_2,m)^{4^{n-1}}$.  We can
now define a series of $n$ order composite pulses P$n$ as
\ba
\mbox{P}0&=&\M{0}{\theta}\\ 
\mbox{P}2&=&\M{\phi_1}{2\pi}\M{-\phi_1}{4\pi}\M{\phi_1}{2\pi}\mbox{P}0\\
\mbox{P}{2j}&=&S_j(\phi_j,-\phi_j,2)\mbox{P}0
\ea
where $\phi_j=\cos^{-1} -\frac{\theta}{8\pi f_j}$ and
$f_j=(2^{(2j-1)}-2)f_{j-1}$ when $f_1=1$.  $\mbox{P}2$ is exactly the
passband sequence PB1 described by Wimperis \cite{Wimperis:94}.
Fig.~\ref{fig:signal} compares the performance of these high-order
passband pulse sequences.


Wimperis also proposes a similar broadband sequence,
BB1=$S_1(\phi_{B1},3\phi_{B1},1)\mbox{P}0$ where $\phi_{B1}=\cos^{-1}
\left(-\frac{\theta}{4\pi}\right)$. The broadband sequence corrects
over a wider range of $\e$ by minimizing the first order commutator
and thus the leading order errors.  Furthermore, although BB1 and PB1
appear different when written as imperfect rotations, a transformation
to true rotations shows that they have the same form,
\ba
  \mbox{PB}1&=&\M{\phi_1}{2\pi}\M{-\phi_1}{4\pi}\M{\phi_1}{2\pi}P0\non
\\
  &=&\R{\phi_1}{2\pi\e}\R{-\phi_1}{4\pi\e}\R{\phi_1}{2\pi\e}\mbox{P}0
\non\\
 \mbox{BB}1&=&\M{\phi_{B1}}{\pi}\M{3\phi_{B1}}{2\pi}\M{\phi_{B1}}{\pi}\mbox{P}0
\\
  &=&\R{\phi_{B1}}{\pi\e}\R{-\phi_{B1}}{2\pi\e}\R{\phi_{B1}}{\pi\e}\mbox{P}0.
\ea
This ``toggled'' frame suggests a way to create higher-order
broadband pulses. One simply takes a higher-order passband sequence
and replaces each element $S_1(\phi_j,-\phi_j,m)$ with
$S_1(\phi_{Bj},-\phi_{Bj}+4\phi_{Bj}(m/2\mod 2),m/2)$ where $\phi_{Bj}$
satisfies the condition $\cos (\phi_{Bj})=2\cos (\phi_j)$. Applying
this to P$n$ creates a family of broadband composite pulses, B$n$.

Similar extensions allow creation of another kind of composite pulse
(useful, for example, in magnetic resonance imaging), which {\em
increase} error so as to perform the desired operation for only a small window
of error.  Such ``narrowband'' pulse sequences N$n$ may be obtained
starting with a passband sequence, P$n$, and dividing the angles of
the corrective pulses by $2$. These higher-order narrowband pulses may
be compared with the Wimperis sequence NB$1$ \cite{Wimperis:94}, as
shown in \fig{signal}.

The SK method yields a third set of $n$th order composite pulses,
SK$n$, and for concreteness, we present an explicit formulation of
this method.  It is convenient to let $U_{nX}(a)=I-ia^n\frac{X}{2}\e^n
+O(\e^{n+1})$, such that one can then generate
$U_{nZ}(a)=\M{90}{-\pi/2}U_{nX}(a)\M{90}{\pi/2}$ and
$U_{nY}(a)=\M{45}{\pi/2}U_{nX}(a)\M{45}{-\pi/2}$ \endnote{The optimal
way to generate $U_{nY}$ is to shift the phases, $\phi$, of the
underlying $\M{\phi}{\theta}$ that generate $U_{nx}$ by 90 degrees.}.
Using the first-order rotations
\be
	U_{1X}(a)=\M{\phi}{2\pi\ceil{\frac{a}{4\pi}}}
		  \M{-\phi}{2\pi \ceil{\frac{a}{4\pi}}}
\,,
\ee 
where $\phi=\cos^{-1} (a/(4\pi\ceil{\frac{a}{4\pi}}))$, as described
above, we may recursively construct $U_{nX}(a) =
U_{\floor{n/2}Y}(a)U_{\ceil{n/2}Z}(a)U_{\floor{n/2}Y}(-a)U_{\ceil{n/2}Z}(-a)$,
for any $n>1$ and any $a$.

With these definitions, the first order SK composite pulse for
$\Rn{n}{0}{\theta}$ is simply 
\ba
	\mbox{SK}1=U_{1X}(\theta)\M{0}{\theta}
		=\R{0}{\theta}-i\frac{A_2}{2}\e^2 +O(\e^3) 
\,.
\ea
From the $2\times 2$ matrix $A_2$, we can then calculate the norm
$\|A_2\|$ and the planar rotation $R_{A_2}$ that performs $R_{A_2}
(-A_2) R_{A_2}^{-1}=\|A_2\| X$.  The second order SK composite pulse
is then
\ba
    \mbox{SK}2 &=& M_{A_2}^{-1} U_{2X}(\|A_2\|^{1/2}) M_{A_2} {\rm SK}1
\\
	&=& \R{0}{\theta}-i\frac{A_3}{2}\e^3+O(\e^4)
\ea
where $M_{A_2}$ is the imperfect rotation corresponding to the perfect
rotation $R_{A_2}$. 

The $n$th order SK composite pulse family is thus
\ba
	\mbox{SK}n &=& M_{A_n}^{-1} U_{nX}(\|A_n\|^{1/n}) M_{A_n}
				\, {\rm SK}(n-1)
\\ 
	&=&\R{0}{\theta}-i\frac{A_{n+1}}{2}\e^{n+1}+O(\e^{n+2})
\,. 
\ea 
A nice feature of the SK method is that when given a composite pulse
of order $n$ described by any method one can compose a pulse of order
$n+1$.  The ``pure'' SK method SK$n$ is outperformed in terms of both
error reduction and pulse number by the TS method B$n$ for $n\leq 4$.
Therefore, we apply the SK method for orders $n>4$ using B$4$ as our
base composite pulse.  We label these pulses SB$n$.
  
\begin{figure}
\begin{center}
\includegraphics[width=8cm]{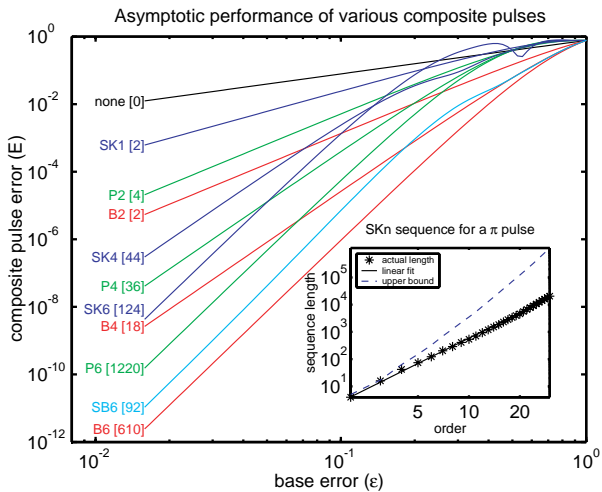}
\end{center}
\caption{\label{fig:effective-error} Composite pulse error, $E$, as a
function of base error, $\epsilon$ for a variety of composite pulse
sequences.  P$n$, B$n$, SK$n$ and SB$n$ are the $n$th order passband,
broadband, SK, and combined B4-SK sequences, respectively.  The number
in the brackets refers to the number of imperfect $2\pi$ rotations in
the correction sequence.  Note how pulses of the same order (such as
P6, B6, SK6, SB6) have the same slope (asymptotic scaling) for low
values of $\epsilon$, but can have widely varying performance when
$\epsilon$ is large.
The inset plots the scaling of this sequence length with order $n$ for
SK$n$ (SB$n$ is very similar) for $n\leq 30$ and compares it with the
upper bound obtained with numerical methods.}
%
\label{fig:logperf}
\end{figure}

\noindent {\bf Performance and efficiency.}  Two important issues with
composite pulses are the actual amount of error reduction as a
function of pulse error, and the time required to achieve a desired
amount of error reduction.  These performance metrics are shown in
\fig{effective-error}, comparing the SK, broadband, and passband
composite pulses for varying error $\e$, and $\phi=0$, and using as
the composite pulse error
$E=\|\R{\phi}{\theta}-\Rn{n}{\phi}{\theta}\|$.  We find that for
practical values of error reduction, $n<30$, the number of $\pi$
pulses required to reduce error to $O(\e^n)$ grows as $\sim n^{3.09}$,
which is close to the lower bound of $\sim n^3$ which can be
analytically derived\cite{long-paper}. In contrast, the TS sequence B$n$ requires $O(\exp (n^2))$ pulses.

For a wide range of base errors $\epsilon$, the TS formulation
out-performs the SK method in achieving a low composite pulse error,
$E$.  The recursive nature of the TS methods builds off elements that
remove lower order errors, resulting in a rapid increase of pulse
number and a monotonic decrease in effective error at every order for
any value of the base error.  However, the SK approach is superior to
the TS method for applications requiring incredible precision, $E\leq
10^{-12}$, from relatively precise controls, $\e<10^{-2}$.





The SK and TS pulse sequences presented here are conceptually simple
but may not be optimal.  Integrating ideas from both methods, we can
develop new families of composite pulses. As an example, the SK method
relies on cancellation of error order by order by building up
sequences of $2\pi$ pulses.  However, there is no reason that the
basic unit should be a single pulse. Instead, one can build a sequence
from TS (B2) style pulse triplets,
$G(\phi_1)=S_1(\phi_1,3\phi_1,1)$. By using an additional symmetry
that the $\tr\left( YG(-\phi_1)G(\phi_1)\right)=0$, the leading order
error is guaranteed to be proportional to $X$ at the cost of doubling
the pulse sequence. The resulting pulses are of length $\exp(n)$
(compared to $\exp(n^2)$ for TS), broadband compared to SK sequences,
and described in detail in \cite{long-paper}.

\noindent{\bf Conclusions.}  We have presented a set of tools that allows one to generate arbitrarily accurate composite pulse sequences for systematic, but unknown, error. As an example, we have constructed explicit composite pulse sequences for errors in rotation angle. These can be constructed with $O(n^3)$ pulses, for
$n\lesssim 30$.  For high-precision applications such as quantum
computation, these pulses allow one to perform accurate operations
even with large errors.  Practically, the B4 and B2=BB1 pulse
sequences seem most useful, depending on the magnitude of error.

While we have focused on composite pulse sequences for
rotation errors, we emphasize that these methods also apply to correcting systematic errors
in control phase and frequency\cite{long-paper}. For example, a frequency error can be represented for an expected rotation $R_0(\theta)$ as $M^\prime_0(\theta)=\exp\left(-i(\theta/2 X +|\theta/2|\delta Z)\right).$ Note that $M^\prime_0(\theta/2)M^\prime_0(-\theta)M^\prime_0(\theta/2)$ yields to first order in $\delta$ the phase shift, $U_{1Z}(2\theta\delta)$. Starting with any fully compensating composite pulse sequence that corrects frequency errors to order $\delta^2$, e.g. CORPSE\cite{Cummins00a}, and the basic operation $U_{1Z}(2\theta\delta)$, one can then apply the SK technique to create a pulse sequence of $O(\delta^n)$\cite{long-paper}.

Furthermore, the TS
and SK approaches can be extended to any set of operations that has a
subgroup isomorphic to rotations of a spin.  For example, Jones has
used this isomorphism to create reliable two qubit gates based on an
Ising interaction to accuracy $O(\e^3)$ \cite{Jones02}.  Similarly,
the techniques outlined here can immediately be applied to gain
arbitrary accuracy multi-qubit gates.  Interestingly, the TS formula
can be directly applied to any set of operations, if the operations
suffer from proportional systematic timing errors.  Therefore, this
control method could also be applied to classical systems.

\begin{acknowledgments}

We are grateful to A. Childs and R. Cleve for stimulating discussions.  AWH was supported by the NSA and ARDA under ARO contract number
DAAD19-01-1-06.

\end{acknowledgments}


\bibliographystyle{apsrev}

\end{document}